\def\imat {\imath}

\documentstyle[twocolumn,floats,epsf,ifthen,aps,prb]{revtex} 

\begin{document} 

 

\title{Continuous measurement of entangled qubits} 

\author{Alexander N. Korotkov} 
\address{ 
Department of Electrical Engineering, University of California, 
Riverside, CA 92521-0204. 
} 
\date{\today} 
 
\maketitle 
 
\begin{abstract} 
We have developed Bayesian formalism to describe the process of continuous 
measurement of entangled qubits. We start with the case of two qubits 
and then generalize it to an arbitrary number of qubits. 
\end{abstract} 
 
\narrowtext 
 
\vspace{0.6cm}

        The problem of quantum measurement of a qubit (two-level system) 
received renewed attention recently in relation to its importance for
quantum computing. The case of sufficiently fast (instantaneous)
measurement can be readily described by ``orthodox'' collapse 
postulate, \cite{Neumann} and this is the case assumed at present by
all quantum algorithms. However, in practice, especially for solid-state
qubits, the act of measurement is not instantaneous. Because of typically
low coupling between a solid-state qubit and a detector, it takes 
a considerable time before the qubit state is completely destroyed by 
the act of measurement. Correspondingly, because of fundamentally
unavoidable noise of the detector, the information about the state
of measured qubit is available not immediately, but only after some
time sufficient to get an acceptably large signal-to-noise ratio. 
It is important that the timescale of measurement (and collapse) process 
may be comparable to the timescale of ``free'' qubit evolution (e.g., due to 
Rabi oscillations) or to the duration of the detector on-off operation 
sequence. (For example, if the detector is switched off when signal-to 
noise ratio is still on the order of unity, the measurement is only 
partially completed.)

        So, for practical needs we should be able to describe the measurement 
of a solid-state qubit as a continuous process. The formalism suitable
to describe 
a continuous measurement of an {\it ensemble} of qubits has been developed 
two decades ago \cite{Caldeira} (for its use in quantum computing 
problems see, e.g., Ref.\ \cite{Makhlin}). In contrast, the formalism
describing the process of measurement of
a {\it single} qubit have been presented only recently 
\cite{Kor-99,Kor-01,Goan} and is still in the stage of active development. 
(In fact, it can be considered as 
a direct continuation of the well-developed field of selective or conditional
quantum measurements -- see, e.g., Refs.\ 
\cite{Davies,Carmichael,Mensky,Caves,Gagen,Presilla} and references in Ref.\ 
\cite{Kor-01}). This formalism is called Bayesian (because of essential
role of the Bayes formula\cite{Borel}) 
and combines advantages of the ``orthodox'' approach \cite{Neumann} 
(ability to treat single quantum systems) and the Leggett's approach 
\cite{Caldeira} (ability to treat continuous measurement). 

        The Bayesian approach has been applied so far only to the continuous
measurement of a single qubit.\cite{Kor-99,Kor-01,Goan,Kor-sp,Kor-exp,Ruskov} 
In this paper we apply it to derive the equations describing continuous 
measurement of entangled qubits.

\begin{figure} [t] 
\centerline{
\epsfxsize=2.2in 
\vspace{0.1cm}
\epsfbox{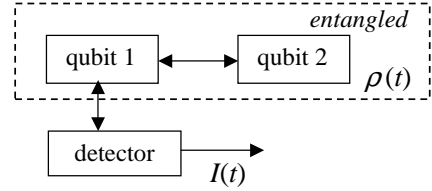}
} 
\vspace{0.3cm} 
\caption{Schematic of two entangled qubits, one of which is continuously 
measured by a detector. The noisy detector output $I(t)$ is used 
to monitor the evolution of the double-qubit density matrix $\rho (t)$. 
 }
\label{Fig1}\end{figure}

        Let us consider first the case of two entangled qubits, one of which 
is continuously measured by a detector (Fig.\ \ref{Fig1}). 
As a main example, we consider qubits 
made of double quantum dots while detector is a quantum point contact 
(realizations based on single-electron transistors and SQUIDs are also 
possible -- see Ref.\ \cite{Kor-01}). 
 Let us denote 4 basis vectors characterizing the state of two qubits as 
$| 1\rangle \equiv\, \mid \uparrow \uparrow \rangle$,
$| 2\rangle \equiv\, \mid \uparrow \downarrow  \rangle$, 
$| 3\rangle \equiv\, \mid \downarrow \uparrow  \rangle$,
and $| 4\rangle \equiv\, \mid \downarrow \downarrow \rangle$. 
(The basis for the first qubit is determined by its interaction with the
detector, while for the second qubit the basis is arbitrary.) 
The qubits can interact with each 
other as well as be noninteracting (the entanglement can be a result of 
previous interaction). The free evolution of qubits is described by the 
Schr\"odinger
equation $d\Psi /dt=(-\imat/ \hbar )H_{qb}\Psi$, where $H_{qb}$ is the 
Hamiltonian of qubits only (not including interaction with the detector). 
Correspondingly, in the case without detector the density matrix 
$\rho$ of double-qubit system evolves as
$d\rho /dt=(-\imat / \hbar )[H_{qb}, \rho ]$. 

	The detector output is characterized by two dc currents, 
$I_\uparrow$ and 
$I_\downarrow$, corresponding to two states of the first qubit, and the
frequency-independent spectral density $S$ of the detector noise. As usual 
\cite{Kor-01} we assume weakly responding (linear) detector, 
$|\Delta I|\ll I_0$, where $\Delta I\equiv I_\uparrow -I_\downarrow$ 
and $ I_0 \equiv (I_\uparrow +I_\downarrow )/2$, to neglect individual
electrons passing through the detector and  consider 
the detector current $I(t)$ as a continuous function of time. For the 
same purpose we assume that the timescale $e/I_0$ (where $e$ is the
electron charge) is much shorter than other timescales in the problem
(due to collapse, dephasing, and free evolution of qubits). 

        Let us start with the simplest case when qubits are ``frozen'', 
$H_{qb}=0$ (so all the evolution is due to the measurement only), 
initial state of qubits is pure, and the detector is ideal
(for example, quantum point contact at low temperature is an ideal 
detector,\cite{Kor-01} as well as single-electron transistor well inside
the cotunneling range \cite{SET-ideal}). 
We can always represent initial pure state 
as $\Psi = \alpha \mid \uparrow \rangle \otimes (a_\uparrow 
\mid \uparrow \rangle + b_\uparrow \mid \downarrow \rangle )
+\beta \mid \downarrow \rangle \otimes (a_\downarrow 
\mid \uparrow \rangle + b_\downarrow \mid \downarrow \rangle )$,
where the states of the second qubit are normalized,  
$|a_\uparrow |^2 +|b_\uparrow |^2 = |a_\downarrow |^2 +|b_\downarrow |^2 
=1$, and consequently $|\alpha |^2+|\beta |^2 =1$. Since the detector
is not coupled to the second qubit, the evolution due to measurement
affects only the factors $\alpha$ and $\beta$, which can be calculated 
using single-qubit Bayesian result \cite{Kor-01} (the overall wavefunction 
phase is of course not important): 
        \begin{eqnarray}
\frac{\alpha (\tau )}{\alpha (0)} = \left[
\frac{P_\uparrow (\tau )}{|\alpha (0)|^2 P_\uparrow (\tau )  +
|\beta (0)|^2 P_\downarrow (\tau ) } \right] ^{1/2} ,
        \label{alpha}   \\
\frac{\beta (\tau )}{\beta (0)} = \left[
\frac{P_\downarrow (\tau )}{|\alpha (0)|^2 P_\uparrow (\tau )  +
|\beta (0)|^2 P_\downarrow (\tau ) } \right] ^{1/2} ,
        \label{beta}    \end{eqnarray}
where $P_\uparrow (\tau )$ and $P_\downarrow (\tau )$ characterize 
the conditional probabilities (for the first qubit in $\mid \uparrow\rangle$
and $\mid \downarrow\rangle$  states) of getting a particular realization of 
the detector output \cite{realization} $I(t)$: 
        \begin{eqnarray}
&& P_\uparrow (\tau ) = (2\pi D)^{-1/2} \exp \left[
 -\left(\overline{I}(\tau )-I_\uparrow \right)^2/2D \right],  
        \label{Pup}\\
&& P_\downarrow (\tau ) = (2\pi D)^{-1/2} \exp \left[ 
-\left(\overline{I}(\tau )-I_\downarrow \right)^2/2D \right],  
        \label{Pdown}\\
&& \overline{I}(\tau ) \equiv \tau^{-1}\int_0^\tau I(t) dt, \,\,\,\, 
D\equiv S/2\tau .
        \label{Iaver}\end{eqnarray} 
In the language of double-qubit density matrix the evolution described by
Eqs.\ (\ref{alpha}) and (\ref{beta}) can be rewritten as 
        \begin{eqnarray}
&& \frac{\rho_{11}(\tau )}{\rho_{11}(0)}=\frac{\rho_{22}(\tau 
)}{\rho_{22}(0)}=
\frac{\rho_{12}(\tau )}{\rho_{12}(0)}=
\frac{P_\uparrow (\tau )}{\rho_\uparrow P_\uparrow (\tau )  +
\rho_\downarrow P_\downarrow (\tau ) } , 
        \label{br11} \\
&& \frac{\rho_{33}(\tau )}{\rho_{33}(0)}=\frac{\rho_{44}(\tau 
)}{\rho_{44}(0)}=
\frac{\rho_{34}(\tau )}{\rho_{34}(0)}=
\frac{P_\downarrow (\tau )}{\rho_\uparrow P_\uparrow (\tau ) +
\rho_\downarrow P_\downarrow (\tau ) } ,
        \label{br33} \\
&& \frac{\rho_{13}(\tau )}{\rho_{13}(0)}=\frac{\rho_{14}(\tau 
)}{\rho_{14}(0)}=
\frac{\rho_{23}(\tau )}{\rho_{23}(0)}=\frac{\rho_{24}(\tau )}{\rho_{24}(0)}
        \nonumber \\
&& \hspace{1.3cm} = \frac{[P_\uparrow (\tau ) P_\downarrow (\tau )]^{1/2} }
{\rho_\uparrow P_\uparrow (\tau )  + \rho_\downarrow P_\downarrow (\tau ) } , 
        \label{br13}\end{eqnarray}
where $\rho_\uparrow \equiv \rho_{11}(0)+\rho_{22}(0)$ and 
$\rho_\downarrow \equiv \rho_{33}(0)+\rho_{44}(0)$ correspond to initial
probabilities to find the first qubit in $\mid\uparrow \rangle$ and 
$\mid \downarrow  \rangle$ states.

        If the initial state $\rho (0)$ is not pure, its evolution can be 
calculated in the following way. Let us represent $\rho (0)$ as
        \begin{equation}
\rho (0) = \sum_s p_s(0) \rho_s(0),
        \end{equation}
where $p_s$ is the classical probability of a pure state $|s\rangle$,
$\rho_s$ is its density matrix, and the sum is over a necessary number 
of pure states. (Of course, such representation is not unique in general.) 
To calculate $\rho (\tau )$ we can apply ``double Bayes'' procedure: 
classical Bayes theorem to obtain probabilities $p_s(\tau )$, 
        \begin{equation}
p_s(\tau ) = \frac{p_s(0)[\rho_{s,\uparrow}P_{\uparrow}(\tau )+
\rho_{s,\downarrow}P_{\downarrow}(\tau )]}
{\sum_r p_r(0) [\rho_{r,\uparrow}P_{\uparrow}(\tau )+
\rho_{r,\downarrow}P_{\downarrow}(\tau )]} ,
        \end{equation}
and the quantum Bayesian result [Eqs.\ (\ref{br11})--(\ref{br13})] 
to calculate each $\rho_s(\tau )$. It is easy to show that the resulting 
evolution of $\rho (\tau )=\sum_s p_s(\tau )\rho_s(\tau )$ satisfies 
Eqs.\ (\ref{br11})--(\ref{br13}), which therefore are valid for
arbitrary mixed states as well. Notice that Eq.\ (\ref{br13}) has 
an obvious interpretation as the conservation of the ``degree of purity''
similar to the one-qubit case. \cite{Kor-01} 

        Besides the derivation of Eqs.\ (\ref{br11})--(\ref{br13}) 
using one-qubit Bayesian result (as above), they can also be obtained
(in the case of pure states) directly using the ``quantum Bayes theorem'', 
\cite{Gardiner} which says that the classical Bayes formula \cite{Borel} 
is applicable not only to the probabilities described by the diagonal 
matrix elements (that is obvious because of the correspondence principle), 
but also applicable to the wavefunction amplitudes. 
Besides that, Eqs.\ (\ref{br11})--(\ref{br13}) can 
be easily derived ``microscopically'' in the case of a low-transparency 
quantum point contact at zero temperature. In this case, solving the 
Schr\"odinger 
equation for the qubits coupled to the detector (for the model see Refs.\ 
\cite{Gurvitz} and \cite{Kor-01}) one can obtain the following Bloch 
equations \cite{Gurvitz} for the density matrix $\tilde{\rho}_{ij}^n$ 
which contains index $n$ corresponding to the number of electrons 
passed through the detector: 
        \begin{eqnarray}
&& d\tilde\rho_{11}^n/dt = -(I_\uparrow /e) \tilde\rho_{11}^n 
   + (I_\uparrow /e) \tilde\rho_{11}^{n-1} ,
        \\
&& d\tilde\rho_{33}^n/dt = -(I_\downarrow /e) \tilde\rho_{33}^n 
   + (I_\downarrow /e) \tilde\rho_{33}^{n-1} ,
        \\
&& d\tilde\rho_{12}^n/dt = -(I_\uparrow /e) \tilde\rho_{12}^n 
   + (I_\uparrow /e) \tilde\rho_{12}^{n-1} ,
        \\
&& d\tilde\rho_{13}^n/dt = -(I_0 /e) \tilde\rho_{13}^n 
   + (\sqrt{I_\uparrow I_\downarrow} /e) \tilde\rho_{13}^{n-1} ,
        \\
&& d\tilde\rho_{14}^n/dt = -(I_0 /e) \tilde\rho_{14}^n 
   + (\sqrt{I_\uparrow I_\downarrow} /e) \tilde\rho_{14}^{n-1} .
        \end{eqnarray}
The equations for other components of $\tilde\rho_{ij}^n$ are similar 
and can be obtained by the substitutions:  
$\tilde\rho_{11}^n \rightarrow \tilde\rho_{22}^n$, 
$\tilde\rho_{33}^n \rightarrow \tilde\rho_{44}^n$, 
$\tilde\rho_{33}^n \rightarrow \tilde\rho_{34}^n$, 
$\tilde\rho_{13}^n \rightarrow \tilde\rho_{23}^n$, 
and $\tilde\rho_{13}^n \rightarrow \tilde\rho_{24}^n$. 
Solving these equations and collapsing the number $n$ at time $\tau$ 
(measuring the charge passed through the detector and obtaining,
for example, charge $me$), 
     \begin{eqnarray}
&& \tilde\rho_{ij}^n (\tau +0) = \delta_{nm}\, \rho_{ij} (\tau +0), 
        \\
&& \rho_{ij} (\tau +0) = \frac{\tilde\rho_{ij}^m (\tau -0)}
{\sum_k\tilde\rho_{kk}^m (\tau -0)},
        \end{eqnarray}
one reproduces Eqs.\ (\ref{br11})--(\ref{br13}). 

        Now let us take into account finite detector ideality
(efficiency), $\eta \leq 1$, 
where in one-qubit case \cite{Kor-01} $\eta \equiv (\Delta I)^2/4S\Gamma$ 
is the ratio of the ``information acquisition rate'' \cite{infacq}
$(\Delta I)^2/4S$ and the ensemble dephasing rate $\Gamma$. 
Similar to the derivation of Ref.\ \cite{Kor-99}, let us consider first 
the case of a detector with neglected output (which is equivalent to
``pure environment''). Then, averaging Eqs.\ (\ref{br11})--(\ref{br13})
over the probability distribution  $\rho_\uparrow P_\uparrow (\tau )
+\rho_\downarrow P_\downarrow (\tau )$ of $\overline I (\tau)$ 
[see Eqs.\ (\ref{Pup})--(\ref{Iaver})], we get the following: 
the right-hand side of Eqs.\ (\ref{br11}) and (\ref{br33}) becomes
unity (which means that $\rho_{11}$, $\rho_{22}$, $\rho_{12}$, 
$\rho_{33}$, $\rho_{44}$,
and $\rho_{34}$ do not change on average), while the right-hand side 
of Eq.\ (\ref{br13}) is replaced by $\exp [-(\Delta I)^2/4S]$
(which means that $\rho_{13}$, $\rho_{14}$, $\rho_{23}$, and 
$\rho_{24}$ decay
on average with the rate $(\Delta I)^2/4S$). Similar to the one-qubit case,
we can regard a nonideal detector as two detectors ``in parallel'', 
\cite{nonideal} neglecting the output of the second detector. 
In this way we obtain the following result for a nonideal detector: 
Eqs.\ (\ref{br11}) and (\ref{br33}) remain valid, while Eq.\
(\ref{br13}) should be replaced by 
        \begin{eqnarray}
\frac{\rho_{13}(\tau )}{\rho_{13}(0)}=\frac{\rho_{14}(\tau )}{\rho_{14}(0)}=
\frac{\rho_{23}(\tau )}{\rho_{23}(0)}=\frac{\rho_{24}(\tau )}{\rho_{24}(0)}
        \nonumber\\
= \frac{[P_\uparrow (\tau ) P_\downarrow (\tau )]^{1/2} }
{\rho_\uparrow P_\uparrow (\tau )  + \rho_\downarrow P_\downarrow (\tau ) } 
\, \exp (-\gamma \tau), 
        \label{nondiag}\end{eqnarray}
where $\gamma = (\eta ^{-1}-1)(\Delta I)^2/4S$. 

        Notice that since the second qubit is not coupled to the detector, 
the state of the second qubit changes only due to its entanglement 
with the first qubit. 
In particular, in the case of no initial entanglement (when $\rho (0)$ 
can be represented as a direct product), the state remains disentangled, 
the second qubit density matrix does not change, and Eqs.\ 
(\ref{br11})--(\ref{br13}) reduce to the Bayesian result 
for the first qubit. 

     If the qubits are not frozen, $H_{qb}\neq 0$, the evolution due to 
$H_{qb}$ should be added to the evolution due to measurement.
In differential form [we use Stratonovich representation, \cite{Kor-01} 
so we take 
usual derivatives of Eqs.\ (\ref{br11}), (\ref{br33}), and (\ref{nondiag})] 
we get the following Bayesian equations: 
        \begin{eqnarray}
&&\dot{\rho}_{11} = \frac{-\imat}{\hbar} [H_{qb},\rho ]_{11} +
\rho_{11} (\rho_{33}+\rho_{44}) \frac{2\Delta I}{S} \left( I(t)-I_0 \right) ,
        \label{r11} \\
&&\dot{\rho}_{33} = \frac{-\imat}{\hbar} [H_{qb},\rho ]_{33} - 
\rho_{33} (\rho_{11}+\rho_{22}) \frac{2\Delta I}{S} \left( I(t)-I_0 \right) ,
        \label{r33} \\
&&\dot{\rho}_{12} = \frac{-\imat}{\hbar} [H_{qb},\rho ]_{12} +
\rho_{12} (\rho_{33}+\rho_{44}) \frac{2\Delta I}{S} \left( I(t)-I_0 \right) ,
        \label{r12} \\
&&\dot{\rho}_{34} = \frac{-\imat}{\hbar} [H_{qb},\rho ]_{34} - 
\rho_{34} (\rho_{11}+\rho_{22}) \frac{2\Delta I}{S} \left( I(t)-I_0 \right) ,
        \label{r34} \\
&& \dot{\rho}_{13} = \frac{-\imat}{\hbar} [H_{qb},\rho ]_{13} 
-\rho_{13} (\rho_{11}+\rho_{22}-\rho_{33}-\rho_{44}) \frac{\Delta I}{S} 
        \nonumber \\ 
&& \hspace{1.0cm} 
\times \left( I(t)-I_0 \right) - \gamma \rho_{13} , 
        \label{r13}     \\
&& \dot{\rho}_{14} = \frac{-\imat}{\hbar} [H_{qb},\rho ]_{14} 
-\rho_{14} (\rho_{11}+\rho_{22}-\rho_{33}-\rho_{44}) \frac{\Delta I}{S} 
        \nonumber \\ 
&& \hspace{1.0cm} 
\times \left( I(t)-I_0 \right) - \gamma \rho_{14} . 
        \label{r14}\end{eqnarray}
The equations for remaining components can be obtained from
Eq.\ (\ref{r11}) by substitution $\{11\}\rightarrow \{22\}$, from 
Eq.\ (\ref{r33}) by substitution $\{33\}\rightarrow \{44\}$, and from 
Eq.\ (\ref{r13}) by substitutions $\{13\}\rightarrow \{23\}$ 
and $\{13\}\rightarrow \{24\}$.  

        These equations allow us to monitor the evolution of the
double-qubit density matrix if we know the initial state $\rho (0)$ 
(for example, we have prepared qubits ourselves) 
and we know the detector output $I(t)$ from an experiment. [To emphasize 
the noisy nature of
$I(t)$ we show this time dependence in Eqs.\ (\ref{r11})--(\ref{r14}) 
explicitly, while the time dependence of the density matrix $\rho$ is not 
shown explicitly.] To simulate the measurement process numerically, 
we need (similar to Ref.\ \cite{Kor-01}) to complement these 
equations  by the formula
        \begin{equation}
I(t)-I_0=\frac{\Delta I}{2} (\rho_{11}+\rho_{22}-\rho_{33}-\rho_{44})
+\xi (t),
        \label{xi} \end{equation}
where $\xi (t)$ is a zero-correlated (``white'') random process with zero
average and the same spectral density as the detector noise, $S_\xi =S$. 
[Eq.\ (\ref{xi}) is derived from the probability distribution 
$\rho_\uparrow P_\uparrow 
(\tau ) +\rho_\downarrow P_\downarrow (\tau )$ for the average current 
$\overline I(\tau )$ at sufficiently small $\tau$, so that evolution
due to $H_{qb}$ can be neglected.]

\begin{figure} [t] 
\centerline{
\epsfxsize=2.7in 
\vspace{0.1cm}
\epsfbox{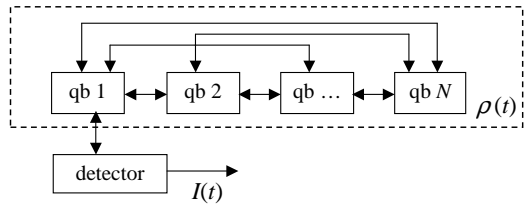}
} 
\vspace{0.3cm} 
\caption{Schematic of $N$ entangled qubits, one of which is continuously 
measured by a detector. Bayesian formalism allows us to monitor $N$-qubit
density matrix $\rho (t)$, using detector output $I(t)$.  
 }
\label{Fig2}\end{figure}

\vspace{0.3cm}

        The generalization of Eqs.\ (\ref{r11})--(\ref{xi}) to the case
of arbitrary number $N$ of entangled qubits, one of which is being 
continuously measured (Fig.\ \ref{Fig2}), is pretty obvious. If both 
basis vectors $i$ and $j$ (from the set of $2^N$ basis vectors) correspond 
to the state $\mid \uparrow \rangle$ of the measured qubit, then the evolution
of the matrix element $\rho_{ij}$ is given by the equation 
        \begin{equation}
\dot{\rho}_{ij} = \frac{-\imat}{\hbar} [H_{qb},\rho ]_{ij} +
\rho_{ij} \, \rho_\downarrow \frac{2\Delta I}{S} \left( I(t)-I_0 \right) .
        \end{equation} 
If both $i$ and $j$ correspond to the state $\mid \downarrow \rangle$ 
of the measured qubit, then 
        \begin{equation}
\dot{\rho}_{ij} = \frac{-\imat}{\hbar} [H_{qb},\rho ]_{ij} - 
\rho_{ij} \, \rho_\uparrow \frac{2\Delta I}{S} \left( I(t)-I_0 \right) .
        \end{equation} 
Finally, if $i$ corresponds to the state $\mid \uparrow \rangle$ while $j$
corresponds to the state $\mid \downarrow  \rangle$, then 
        \begin{equation}
\dot{\rho}_{ij} = \frac{-\imat}{\hbar} [H_{qb},\rho ]_{ij} 
-\rho_{ij} (\rho_\uparrow-\rho_\downarrow) \frac{\Delta I}{S} 
\left( I(t)-I_0 \right) - \gamma \rho_{ij} . 
        \end{equation} 
In these equations $H_{qb}$ is again the Hamiltonian of qubits (without 
detector) while $\rho_\uparrow (t) $ and $\rho_\downarrow (t)$ (now 
time-dependent) are the sums of the diagonal matrix elements of $\rho (t)$, 
corresponding to the states 
$\mid \uparrow  \rangle$ and $\mid \downarrow \rangle$ 
of the measured qubit. Eq.\ (\ref{xi}) should be generalized as 
        \begin{equation}
I(t)-I_0=\frac{\Delta I}{2} \left( \rho_\uparrow (t) -\rho_\downarrow (t) 
\right) +\xi (t). 
        \label{xi-2} \end{equation}

\begin{figure} [t] 
\centerline{
\epsfxsize=2.7in 
\vspace{0.2cm}
\epsfbox{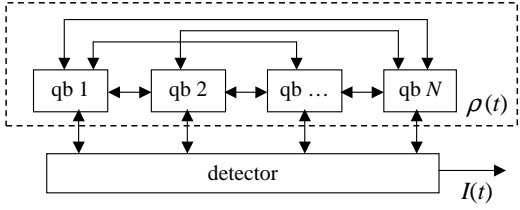}
} 
\vspace{0.3cm} 
\caption{$N$ entangled qubits, continuously measured by a detector, coupled 
to all qubits. 
 }
\label{Fig3}\end{figure}

        Now let us generalize the formalism to the case when the detector
is coupled to all qubits (Fig.\ \ref{Fig3}). Classically, in this case 
there are up to $2^N$  different dc current levels $I_i$, 
corresponding to various combinations of qubit states. Some of these 
levels can coincide, for example, if the detector is not coupled to
some qubits or if some qubits are coupled to the detector equally strong. 
Applying the quantum Bayes theorem in the case of frozen qubits, 
$H_{qb}=0$, and taking into account finite ideality $\eta$ of the detector, 
we obtain the following equations: 
        \begin{eqnarray}
&& \frac{\rho_{ij}(\tau )}{\rho_{ij}(0)}= \frac{\sqrt{P_i(\tau )P_j(\tau )}}
{\sum_k \rho_{kk}(0) P_k(\tau )} \, \exp (-\gamma_{ij}\tau ),
        \label{gen-r}\\ 
&& P_i(\tau )= (2\pi D)^{-1/2} \exp \left[ 
 -\left(\overline{I}(\tau )-I_i \right)^2/2D \right] , 
        \label{gen-p}\\
&& \gamma_{ij}=(\eta^{-1} -1)(I_i-I_j)^2/4S, 
        \label{gen-g}\end{eqnarray}
where $\overline{I}(\tau )$ and $D$ are defined by Eq.\ (\ref{Iaver}),
and the sum in Eq.\ (\ref{gen-r}) is over all $2^N$ basis vectors $k$ 
(the basis is defined by the interaction between the detector and each qubit). 
Correspondingly, the probability distribution of $\overline I (\tau )$ 
is $\sum_i \rho_{ii}(0) P_i(\tau )$. 
Notice that the exponent due to nonideality in Eq.\ (\ref{gen-r}) 
disappears for diagonal 
matrix elements ($i=j$) and also if the classical currents $I_i$ and $I_j$
for two different configurations coincide. This is because $I_i=I_j$
means equal coupling of the detector to the states $i$ and $j$, so
the detector noise cannot destroy the coherence between these states.

        Let us briefly discuss what will happen to Eqs.\ 
(\ref{gen-r})--(\ref{gen-g}) if we relax the assumption of weak detector
response, $|I_i-I_j|\ll (I_i+I_j)/2$. As an example, let us again consider
a low-transparency quantum point contact at zero temperature. In the case 
of moderate or strong response, each electron passed through the detector 
brings a significant information and, correspondingly, changes significantly 
the density matrix $\rho$ of the qubits. Then the language of continuous 
detector 
current is not applicable anymore, and instead of considering average current 
$\overline I (\tau )$ we should count the number $n$ of electrons passed
through the detector during time $\tau$. Equation (\ref{gen-r}) in this case
does not change (except $\gamma_{ij}=0$ since the detector is ideal),
while the Gaussian distribution in Eq.\ (\ref{gen-p}) should be replaced by
the Poissonian distribution: 
$P_i(\tau )= (n!)^{-1}(I_i\tau /e)^{n}\exp (-I_i\tau /e)$.
        It is not easy to introduce nonideality for a detector with finite 
response. If, however, we define $\eta$ in a way \cite{Goan} similar to
optical quantum efficiency as a probability to observe an electron tunneled
through a detector (unfortunately, this definition is hardly justified
in typical solid-state setups), then we can keep the exponential term in Eq.\
(\ref{gen-r}) and should replace Eq.\ (\ref{gen-g}) by 
$\gamma_{ij}=(\eta^{-1}-1) (\sqrt{I_i}-\sqrt{I_j})^2/2e$.

        Returning to the case of weak detector response and continuous 
current, differentiating 
Eq.\ (\ref{gen-r}) over time, and adding the 
free evolution due to $H_{qb}$, we finally obtain the following equation:
        \begin{eqnarray}
&&\dot{\rho}_{ij} = \frac{-\imat}{\hbar} [H_{qb},\rho ]_{ij} + 
\rho_{ij} \, \frac{1}{S} \sum_k \rho_{kk} \left[ 
\left(I(t)-\frac{I_k+I_i}{2}\right) \right. 
        \nonumber \\
&& \hspace{1.0cm} \left. \times  (I_i-I_k)+
\left(I(t)-\frac{I_k+I_j}{2}\right) (I_j-I_k) \right] 
        \nonumber \\ 
&& \hspace{1.0cm} -\gamma_{ij}\rho_{ij} . 
        \label{gen-gen}\end{eqnarray}
Equation (\ref{xi}) in this case is replaced by 
        \begin{equation}
I(t) =\sum_i \rho_{ii} (t) I_i +\xi (t).
        \end{equation}

        Our final generalization is to the case of several detectors,
coupled to $N$ qubits. Each detector has its own set of up to $2^N$
classical current levels. It is important to notice that coupling of
qubits to different detectors can define different sets of basis 
vectors. So, generalization of Eq.\ (\ref{gen-gen}) requires to
sum the terms due to measurement over all detectors, choosing
particular basis for each detector.

        In conclusion, we have developed the Bayesian formalism 
describing continuous measurement of entangled solid-state qubits.
The case of two qubits, one of which is measured by a detector is considered
in detail and then generalized to an arbitrary case. For nonideal
detectors we have assumed the absence of correlation between output
and backaction noises, so the formalism applicable to nonideal detectors 
with such correlation \cite{Kor-01} still has to be developed.
The results of this paper can be experimentally tested.
However, such experiments seem to be still a little beyond the reach of 
the present-day solid-state technology. They could be attempted after
proposed Bayesian experiments with a single solid-state qubit,
in particular, Bell-type experiment. \cite{Kor-exp}

The author thanks R. Ruskov for critical reading of the manuscript. 
The work was supported by NSA and ARDA under ARO grant DAAD19-01-1-0491.



\begin{references}
\bibitem{realization} In the case of frozen qubits the probability
     of a particular output realization $I(t)$ is completely determined 
     by the average current $\overline I(\tau )$ of this realization. 
     This is typically not true, however, in the presence of free evolution,
     $H_{qb}\neq 0$.
\bibitem{infacq} Actually, the average classical information acquisition rate
        (bit/s) is $[(\Delta I)^2/S] \, \rho_\uparrow \rho_\downarrow 
        (\rho_\uparrow -\rho_\downarrow ) \log_2 (\rho_\uparrow /
        \rho_\downarrow)$.
\bibitem{nonideal} In the case of nonideal detector we neglect the 
        correlation between the output noise and the backaction noise. 
        In the case of a significant correlation the formalism is 
        more complicated -- see Ref.\ \protect\cite{Kor-01}.
\end{references}
\end{document}